\documentclass[12pt]{article}
\usepackage{amsfonts}
\usepackage{amsmath,amssymb}
\newtheorem{thm}{Theorem}[section]
\newtheorem{prop}[thm]{Proposition}
\newtheorem{lem}[thm]{Lemma}
\numberwithin{equation}{section}
\def\Tr{{\rm Tr}}

\def\H{{\cal H}}

\def\Det{{\rm Det}}
\def\R{\Bbb R}
\def\C{\Bbb C}
\def\Z{\Bbb Z}
\def\N{\Bbb N}
\def\per{{\rm per\,}}

\def\supp{{\rm supp\,}}
\begin{document}

\title {
    A Random Point Field \\
    related to Bose-Einstein Condensation
             }
\author { H. Tamura \\
 Department of Mathematics, Kanazawa University,\\
          Kanazawa 920-1192, Japan\\
          E-mail : tamurah@kenroku.kanazawa-u.ac.jp
   \and K. R. Ito \\
      Department of Mathematics and Physics, Setsunan University,\\
         Neyagawa, Osaka 572, Japan\\
         E-mail : ito@mpg.setsunan.ac.jp \\
    also at : Division of Mathematics,  \\
            College of Human and Environmental Studies,\\
            Kyoto University, Kyoto 606, Japan\\
         E-mail : ito@kurims.kyoto-u.ac.jp\\
}
\maketitle
\begin{abstract}
The random point field which describes the position distribution of the system
of ideal boson gas in a state of Bose-Einstein condensation is obtained through the
thermodynamic limit.
The resulting point field is given by convolution of two independent 
point fields:
the so called boson process whose generating functional is represented by
inverse of the Fredholm determinant for an operator related to the heat operator
and the point field whose generating functional is represented by a resolvent
of the operator.
The construction of the latter point field in an abstract formulation is also given.
\end{abstract}
\section{Introduction}
In the previous paper \cite{TI}, which we will refer as I,  the authors gave a method
which derives typical kinds of random point fields, the boson point process and the fermion 
point process on $\R^d$,  through the thermodynamic limit from random point fields of 
fixed finite numbers of points in bounded boxes in $\R^d$.
The purpose of the paper is to give the random point field which describes the position 
distribution of the system of ideal boson gas in  a state of Bose-Einstein condensation [BEC] 
as an extension of I.

Let us consider the system of $N$ free bosons in a 
box of finite volume $V$ in $\R^d$ and the quantum statistical mechanical
state for the system of a finite temperature. 
Regarding the square of the absolute value of the wave function
as the distribution function of the positions of $N$ particles together with thermal 
average, we obtain a random point field of $N$ points in the box.
In I, the thermodynamic limit, $N, V \to \infty$ and $N/V \to \rho$, 
of the system for small $\rho$ as well as the system of fermions for
every positive $\rho$ are taken to get
the boson as well as the fermion point processes on $\R^d$ in a simple and straightforward way.
As applications of the approach, the system of para-particles
and the system of composite particles are studied.
The argument is based on the unified formulation of boson/fermion processes of \cite{ST}.
For general references of this field, see e.g. \cite{So} and references cited there in.
 
In this paper, we study the case of large $\rho$ (corresponding to BEC)
which needs technically elaborate analysis for the largest eigenvalue $\tilde g_0(L)$
of the deformed heat operator $\tilde G_L$ in the box of size $L$ and the saddle points
 $z_0 $ and $\tilde z_0$ for complex integrals related to generalized Vere-Jones'
formula \cite{V, ST} more than those in I. 
As the result of the thermodynamic limit, we get a random point field on $\R^d$ which
are given by convolution of two independent point fields:
1. the boson process  whose generating functional is represented by
inverse of the Fredholm determinant for an operator related to the heat operator; 
2. the point field whose generating functional is represented by a resolvent
of the operator.

The paper organized as follows:
In \S 2 the construction of the point field which appears in the resulting point
 fields as the second independent component (see above).
The construction is made in a general framework of random point fields similar
to \cite{ST}, i.e., on the locally compact space of second countability.
\S 3 devoted to the analysis of the thermodynamic limit in $\R^d$.

\section{Abstract formulation of the random point field} 

Let $R$ be a locally compact Hausdorff space with countable basis and $\lambda$ a positive
Radon measure on $R$. 
We regard  $ \lambda $ as a measure on the Borel $\sigma$-algebra ${\cal B}(R)$ 
which assigns finite values for compact sets.
Relatively compact subsets of $R$ will be called bounded.
On $ L^2( R; \lambda) $, we consider a (possibly unbounded) non-negative 
self-adjoint operator $K$ which satisfies:

\medskip

\noindent{\bf Condition K}

\noindent (i) {\sl \;[locally boundedness] For any bounded measurable subset $\Lambda$
 of $R$, the operator $ K^{1/2}\chi_{\Lambda} $ is bounded, where $ \chi_{\Lambda} $ 
denotes the operator multiplying the indicator function }$\chi_{\Lambda} $.

\noindent (ii) $ G = K(1+K)^{-1} $ {\sl has a non-negative integral kernel $ G(x,y) $ which satisfies}
\begin{equation}
       \int_R G(x,y)\lambda(dy) \leqslant 1 \qquad \lambda-a.e. \, x \in R.
\label{G<1}
\end{equation}

\medskip

For a measurable function $ f: R \to [0, \infty) $ with compact support and 
a bounded measurable set $\Lambda $ satisfying $ \Lambda \supset \supp f $, we have
 $ K^{1/2}\sqrt{1-e^{-f}} = K^{1/2}\chi_{\Lambda}\sqrt{1-e^{-f}}$ and hence
that
\begin{equation}
    K_f = \big( K^{1/2}\sqrt{1-e^{-f}}\big)^*K^{1/2}\sqrt{1-e^{-f}}
\label{K_f}
\end{equation}
is a bounded non-negative self-adjoint operator.
Here we regard  $\sqrt{1-e^{-f}}$ the multiplication operator of the function
expressed by the same symbol. 
$Q(R)$ denotes the Polish space of all the locally finite non-negative integer valued
Borel measures on $R$.

\begin{thm}
For $ R, \lambda$ and $K$ which satisfy the above conditions and $\rho >0$,
there exists a unique Borel probability measure $\mu_{K, \rho}$ on $Q(R)$ such that
\begin{equation}
       \int_{Q(R)}e^{-<f,\xi>}d\mu_{K,\rho}(d\xi)
       = \exp\big(-\rho(\sqrt{1-e^{-f}}, [1 + K_f]^{-1}\sqrt{1-e^{-f}})\big)
\label{numpr}
\end{equation}
holds for any non-negative measurable function $f$ on $R$ with compact support,
 where $( \, \cdot \, , \, \cdot \, )$ denotes the inner product of $L^2( R; \lambda)$.
\label{thmG}
\end{thm}

Let us begin with some remarks before proving the theorem.
It follows that $G$ is self-adjoint and $ 0 \leqslant G \leqslant 1$,
where $1$ denotes the identity operator on $L^2( R; \lambda)$.
Without loss of generality, we may assume that the ${\cal B}(R^2)$-measurable 
function $G(x,y)$ satisfies
\[
   \forall x, y \in R : \quad G(x,y) \geqslant 0, \quad G(x,y) =G(y,x)
\]
and
\[
    \forall x \in R: \quad \int_R G(x,y)\,\lambda(dy) \leqslant 1.
\]
 Let us define the functions $G^n(x, y)$ inductively as
\[
     G^1(x,y) = G(x,y), \quad \mbox{and} \quad 
        G^{n+1}(x,y) = \int_R G^n(x, z) G(z,y)\,\lambda(dz) 
        \quad \mbox{for} \quad n\in \N.  
\] 
Then we have
\[
   \forall x, y \in R, \, \forall n \in \N :\quad G^n(x,y) \geqslant 0, \quad G^n(x,y) =G^n(y,x)
\]
and
\[
    \forall x \in R, \, \forall n \in \N: \quad \int_R G^n(x,y)\,\lambda(dy) \leqslant 1.
\]
It is obvious that $G^n(x,y)$ is the integral kernel of the operator $G^n$ 
for any $ n\in \N$.

Put
\[
    K_n = \sum_{k=1}^n G^k \quad \mbox{ and } \quad K_n(x,y) = \sum_{k=1}^n G^k(x,y).
\]
Then $K_n$ is the bounded non-negative self-adjoint operator which has non-negative 
integral kernel $K_n(x,y)$.
The function 
\begin{equation}
K(x,y) = \lim_{n\to\infty}K_n(x,y) = \sum_{k=1}^{\infty} G^k(x,y)
\label{Kxy}
\end{equation}
is well defined, if we admit infinity as its value.

Here we recall the following preliminary facts from functional analysis.
\begin{lem}
{\rm (i)} Let $\H$ be a Hilbert space, ${\cal L}(\H)$  the Banach space of all the bounded 
operators on $\H$ and $\{ A_n\}_{n\in\N}$ a bounded increasing sequence of non-negative 
 self-adjoint operators in ${\cal L}(\H)$.
Then {\rm s}-$\lim_{n\to\infty} A_n$ exists and is a bounded non-negative self-adjoint operator.
 
{\rm (ii)} Suppose that $A_1, \cdots, A_n, \cdots \in {\cal L}(L^2(R;\lambda))$ converge
to $A\in {\cal L}(L^2(R;\lambda))$ strongly, $A_n$ has integral kernel $A_n(x,y)$ for each $n$
and 
\[
       0\leqslant A_n(x,y) \uparrow A(x,y) \qquad \lambda^{\otimes 2}-a.e.\, (x,y) \in R^2.
\]
Then $A$ has $A(x,y)$ as its integral kernel.
\label{GA}
\end{lem}
{\sl Proof }: For (i), see e.g. \cite{RN}.
 
For (ii), let $ f\in L^2(R;\lambda)$. Then $|f| \in  L^2(R;\lambda)$ and 
$(A_n|f|)(x) = \int A_n(x,y)|f(y)|\lambda(dy)$ holds.
Taking the limit $n \to \infty$ (through a subsequence if necessary), we have 
$(A|f|)(x) = \int A(x,y)|f(y)|\lambda(dy) \quad \lambda-a.e.$
 by strong convergence of the operators and the monotone convergence theorem.
The a.e. boundedness of the integral in the righthand side ensures
the identity for $f$ instead of $|f|$ by dominated (instead of monotone) 
convergence theorem.
\hfill $\Box$

Now we have the following proposition.
Here and hereafter, $|| \cdot ||$ and  $|| \cdot ||_T$ stand for
the operator norm and the trace norm for operators, respectively, and
 $|| \cdot ||_p$ for the $L^p$-norm for functions.
\begin{prop}
{\rm(i)} Put $K_{\Lambda} = (K^{1/2}\chi_{\Lambda})^*K^{1/2}\chi_{\Lambda}$ for bounded 
measurable $\Lambda \subset R$.
Then, $K_{\Lambda} $ is a bounded non-negative self-adjoint operator and  has 
$ K_{\Lambda}(x,y) \equiv \chi_{\Lambda}(x)K(x,y)\chi_{\Lambda}(y)$ as its integral kernel.
\begin{equation}
   K_{\Lambda} = \sum_{k=1}^{\infty}\chi_{\Lambda}G^k\chi_{\Lambda}
\label{K_L}
\end{equation}
holds in the sense of strong convergence of operators.

{\rm (ii)} For each $k \in \N$, $H_k = \chi_{\Lambda}G((1-\chi_{\Lambda})G)^{k-1}\chi_{\Lambda}$
is a bounded non-negative self-adjoint operator having non-negative kernel,
denoted by $H_k(x,y)$.
$R_{\Lambda}= \sum_{k=1}^{\infty} H_k$ exists in the strong convergence sense and 
is the bounded non-negative self-adjoint operator having non-negative kernel 
$R_{\Lambda}(x,y)=\sum_{k=1}^{\infty}H_k(x,y)$.

{\rm (iii)} $R_{\Lambda} = K_{\Lambda}(1+K_{\Lambda})^{-1}, \quad ||R_{\Lambda}||<1$.

{\rm (iv)} $(1+K_{\Lambda})^{-1}\chi_{\Lambda} \geqslant 0 \quad a.e.$ holds,
where we regard $\chi_{\Lambda}$ as a function which belongs to $L^2(R;\lambda)$.
\label{GB}
\end{prop}
{\sl Remark : } From (i) of the proposition and the argument above (\ref{K_f}),
it follows that $ K_f = \sqrt{1-e^{-f}}K_{\Lambda}\sqrt{1-e^{-f}} $ and its
kernel is given by $\sqrt{1-e^{-f(x)}}K(x,y)\sqrt{1-e^{-f(y)}} $ for non-negative 
$f$ satisfying supp$\, f \subset \Lambda$. 

\medskip

\noindent{\sl Proof : } (i) Boundedness and self-adjointness of $K_{\Lambda}$ are obvious.

\noindent Using the spectral decomposition 
             $ K = \int_0^{\infty}\lambda\,dE_{\lambda}$, we have
\[
          G = \int_0^{\infty}\frac{\lambda}{1+\lambda}\,dE_{\lambda}.
\]
Hence, 
\[
    ||\sum_{k=1}^n \chi_{\Lambda}G^k\chi_{\Lambda}||
    = \sup_{||\phi||_2=1}\sum_{k=1}^n (\chi_{\Lambda}\phi, G^k\chi_{\Lambda}\phi)
\]
\[
    =\sup_{||\phi||_2=1}\int\sum_{k=1}^n \Big(\frac{\lambda}{1+\lambda}\Big)^k
    \,d(\chi_{\Lambda}\phi, E_{\lambda}\chi_{\Lambda}\phi)
    \leqslant \sup_{||\phi||_2=1}\int\lambda
    \,d(\chi_{\Lambda}\phi, E_{\lambda}\chi_{\Lambda}\phi)
\]
\[
    = \sup_{||\phi||_2=1}||K^{1/2}\chi_{\Lambda}\phi||_2^2
    = ||K_{\Lambda}||.
\]
Since $\chi_{\Lambda}G^k\chi_{\Lambda} \geqslant 0$ holds for every $k \in \N$,
Lemma \ref{GA}(i) yields  the existence of 
 s-$\lim_{n \to\infty}\sum_{k=1}^n\chi_{\Lambda}G^k\chi_{\Lambda}$.
On the other hand, thanks to the monotone convergence theorem, we get
(\ref{K_L}) in the weak sense:
\[
   (\phi, \sum_{k=1}^n\chi_{\Lambda}G^k\chi_{\Lambda}\phi) = 
   \int\sum_{k=1}^n \Big(\frac{\lambda}{1+\lambda}\Big)^k
    \,d(\chi_{\Lambda}\phi, E_{\lambda}\chi_{\Lambda}\phi)
\]
\[
    \longrightarrow \int\lambda \,d(\chi_{\Lambda}\phi, E_{\lambda}\chi_{\Lambda}\phi)
     =(\phi, K_{\Lambda}\phi).
\]
Thus we have (\ref{K_L}) in the strong sense.

\noindent Lemma \ref{GA}(ii) yields the assertion on the kernel of $K_{\Lambda}$.

(ii)  It is obvious that $H_k$ is a bounded non-negative self-adjoint operator
for every $k\in \N$.
From the non-negativity of the kernel of $G^k$, we have the non-negativity of 
the kernel $H_k(x,y)$ and 
\[
     0 \leqslant H_k(x,y) \leqslant \chi_{\Lambda}(x)G^k(x,y)\chi_{\Lambda}(y).
\]
Lemma \ref{GA}(i) and the estimate
\[
      ||\sum_{k=1}^n H_k|| = \sup_{||\phi||_2=1}\sum_{k=1}^n 
       \int_{R^2} \overline{\phi(x)}H_k(x,y)\phi(y) \lambda^{\otimes 2}(dx\,dy)
\]
\[
         \leqslant \sup_{||\phi||_2=1}\sum_{k=1}^n 
       \int_{R^2} |\phi(x)|\chi_{\Lambda}(x)G^k(x,y)\chi_{\Lambda}(y)|\phi(y)| 
             \lambda^{\otimes 2}(dx\,dy)
      \leqslant||\sum_{k=1}^n \chi_{\Lambda}G^k\chi_{\Lambda}|| \leqslant ||K_{\Lambda}||,
\]
we get the existence of the strong limit $R_{\Lambda}$ of $\{\sum_{k=1}^n H_k\}_n$
and its bounded self-adjointness.

\noindent Lemma \ref{GA}(ii) yields the assertion on the kernel of $R_{\Lambda}$.

(iii) From
\[
   \sum_{k=1}^nH_k - \sum_{k=1}^n\chi_{\Lambda}G^k\chi_{\Lambda} =
\sum_{k=1}^n\chi_{\Lambda}G[((1-\chi_{\Lambda})G)^{k-1}
             -G^{k-1}]\chi_{\Lambda}
\]
\[
             = \sum_{k=2}^n\sum_{l=1}^{k-1}\chi_{\Lambda}G
           ((1-\chi_{\Lambda})G)^{k-l-1}(-\chi_{\Lambda}G)
             G^{l-1}\chi_{\Lambda}
\]
\[
            = - \sum_{l=1}^{n-1}\sum_{k=l+1}^n\chi_{\Lambda}G
   ((1-\chi_{\Lambda})G)^{k-l-1}\chi_{\Lambda}\chi_{\Lambda}G^l\chi_{\Lambda}
      = - \sum_{l=1}^{n-1}\sum_{m=1}^{n-l}H_m\chi_{\Lambda}G^l\chi_{\Lambda},
\]
we get the relation
\[
   \sum_{k=1}^nH_k(x,y) - \sum_{k=1}^n\chi_{\Lambda}(x)G^k(x,y)\chi_{\Lambda}(y)
\]
\[
 = - \sum_{l=1}^{n-1}\sum_{m=1}^{n-l}\int_R H_m(x,z)\chi_{\Lambda}(z)
        G^l(z,y)\chi_{\Lambda}(y) \lambda(dz)  \quad  a.e.    
\]
in terms of kernels.  Taking the limit $n\to \infty$, 
we get
\[
           R_{\Lambda} (x,y) - K_{\Lambda} (x,y) = 
                -\int_RR_{\Lambda} (x,z) K_{\Lambda} (z,y) \lambda(dz)
           \qquad \lambda^{\otimes 2}-a.e.(x, y)
\]
by the monotone convergence theorem.
It implies $R_{\Lambda}- K_{\Lambda}= - R_{\Lambda}K_{\Lambda}$ and hence
  $R_{\Lambda}= K_{\Lambda}(1+K_{\Lambda})^{-1}$.
Since $K_{\Lambda}$ is non-negative and bounded, $||R_{\Lambda}||<1$.  

(iv) We may regard $G$ as a contraction operator on $L^{\infty}(R;\lambda)$
because of (\ref{G<1}).
$H_k$ is also contraction on $L^{\infty}(R;\lambda)$ for all $k\in \N$.
Thus  we have
\[
     \sum_{k=1}^{n}(H_k\chi_{\Lambda})(x) \leqslant \sum_{k=1}^{n}(H_k\chi_{\Lambda})(x)
        + \big(\chi_{\Lambda} G((1-\chi_{\Lambda} )G)^{n-1}(1-\chi_{\Lambda} )\big)(x)
\]
\[
     \leqslant \sum_{k=1}^{n-1}(H_k\chi_{\Lambda})(x)
        + \big(\chi_{\Lambda} G((1-\chi_{\Lambda} )G)^{n-2}(1-\chi_{\Lambda} )\big)(x)
     \leqslant \cdots \leqslant (\chi_{\Lambda}G1)(x) \leqslant \chi_{\Lambda}(x),
\]
where non-negativity of the kernel of $G$ and (\ref{G<1}) have been used.
On the other hand, we get 
$\sum_{k=1}^{n}H_k\chi_{\Lambda} \to R_{\Lambda}\chi_{\Lambda} \quad a.e.$
 from (ii) through subsequence if necessary.
Hence
$(1+K_{\Lambda})^{-1}\chi_{\Lambda} = \chi_{\Lambda} - R_{\Lambda}\chi_{\Lambda} \geqslant 0 \quad a.e.$
holds.

\hfill $\Box$

\bigskip

\noindent ({\sl Proof of Theorem \ref{thmG}})

\medskip

\noindent Recall that $K_f = \sqrt{1-e^{-f}}K_{\Lambda}\sqrt{1-e^{-f}}$,
for non-negative measurable $f$ and a bounded measurable set $\Lambda \supset \supp f$.
Since
\[
     (1+(1-e^{-f})K_{\Lambda})\sqrt{1-e^{-f}}(1+K_f)^{-1}\sqrt{1-e^{-f}}
    = 1-e^{-f} = 1-e^{-f}R_{\Lambda}-e^{-f}(1+K_{\Lambda})^{-1}
\]
and
\[
  1+(1-e^{-f})K_{\Lambda} = (1-e^{-f}R_{\Lambda})(1+K_{\Lambda}),
\]
we get
\[
    \sqrt{1-e^{-f}}(1+K_f)^{-1}\sqrt{1-e^{-f}} = (1+K_{\Lambda})^{-1}
    (1-e^{-f}R_{\Lambda})^{-1}(1-e^{-f}R_{\Lambda}-e^{-f}(1+K_{\Lambda})^{-1})
\]
\[
      =(1+K_{\Lambda})^{-1}
   [1 - (1-e^{-f}R_{\Lambda})^{-1}e^{-f}(1+K_{\Lambda})^{-1}]
       =  (1+K_{\Lambda})^{-1} - (1+K_{\Lambda})^{-1}
         \sum_{n=0}^{\infty}(e^{-f}R_{\Lambda})^ne^{-f}(1+K_{\Lambda})^{-1}.
\]
The Neumann expansion in the last step is valid since 
$||e^{-f}R_{\Lambda}|| \leqslant ||R_{\Lambda}|| <1$. 
Hence we have
\[
    -(\sqrt{1-e^{-f}}, [1+K_f]^{-1}\sqrt{1-e^{-f}})
\]
\[
       =  -(\chi_{\Lambda}, (1+K_{\Lambda})^{-1}\chi_{\Lambda})
          + \sum_{l=0}^{\infty}((1+K_{\Lambda})^{-1}\chi_{\Lambda},
           e^{-f}(R_{\Lambda}e^{-f})^l(1+K_{\Lambda})^{-1}\chi_{\Lambda}).
\]
Substituting this identity to the right hand side of (\ref{numpr}), expanding
 the exponential and symmetrizing, we get a expression of the form
\begin{equation}
  \sum_{n=0}^{\infty}\frac{1}{n!}
    \int_{\Lambda^n}\sigma_{\Lambda^n}(x_1, \cdots, x_n)e^{-\sum_{k=1}^n
f(x_k)}
         dx_1\cdots dx_n 
\label{kol}
\end{equation}
with a family of symmetric non-negative
functions $\{\sigma_{\Lambda^n}\}$ for every $ \Lambda \supset$ supp$\, f$.
For the existence of the measure $\mu_{K,\rho}$ on $Q(R)$, it is enough to show
the consistency condition\cite{L}:
\[
         \sigma_{\Lambda^n}(x_1, \cdots, x_n)
                   = \sum_{l=0}^{\infty}\frac{1}{l!}\int_{\Delta^l}
                      \sigma_{(\Lambda\cup\Delta)^{n+l}}
                 (x_1, \cdots, x_n, y_1, \cdots, y_l)\,dy_1\cdots dy_l,
\]
where $\Delta \cap \Lambda = \emptyset $.
This condition can be derived easily from the facts that the right hand side of
(\ref{numpr}) does not depend on $ \Lambda \supset$ supp$\, f$
and that for a given $\Lambda$, $\{\sigma_{\Lambda^n}\}$ in (\ref{kol})
is uniquely determined a.e., since $f$ can be arbitrary non-negative measurable 
function satisfying supp$\, f \subset \Lambda$.

Thus we have proved Theorem \ref{thmG}.
\hfill$\Box$

\section{The Thermodynamic Limit}
In this section, we follow the arguments and the notations of I \S 2.2.
However, let us review them briefly to make the article self-contained.

Consider $ {\cal H}_L = L^2(\Lambda_L) $ on
$ \Lambda_L = [-L/2, L/2]^d $ $\subset {\Bbb R}^d$ for $ d > 2$ with the 
Lebesgue measure on $\Lambda_L$.  
Let $\triangle_L$ be the Laplacian under the periodic boundary condition in 
$ {\cal H}_L $.  
For $k\in {\Bbb Z}^d$, 
$ \varphi_k^{(L)}(x) = L^{-d/2} \exp (i2\pi k\cdot x/L) $ 
is an eigenfunction of $\triangle_L$,  and 
$ \, \{\, \varphi_k^{(L)} \, \}_{k\in {\Bbb Z}^d} \,$ forms an 
complete orthonormal system [CONS] of $ \H_L $.
The operator $G_L = \exp(\beta\triangle_L)$ has the integral kernel
\begin{equation}
     G_L(x,y) = \sum_{k\in {\Bbb Z}^d}e^{-\beta  |2\pi k/L|^2} \varphi_k^{(L)}(x)
     \overline{\varphi_k^{(L)}(y)}, 
\label{GL}
\end{equation}
for $\beta > 0$.  We put $ g_k^{(L)} = \exp (-\beta |2\pi k/L|^2) $ which is the 
eigenvalue of $G_L$ for the eigenfunction 
   $\varphi_k^{(L)}(x)$.
We also need the operator $G = \exp(\beta \triangle)$
 on $L^2({\Bbb R}^d)$ and its integral kernel
\[
    G(x,y) = \int_{ {\Bbb R}^d}\frac{dp}{(2\pi)^d}
    e^{-\beta |p|^2 +ip\cdot(x-y)} 
           = \frac{\exp(-|x-y|^2/4\beta)}{(4\pi \beta)^{d/2}}.
\]
Let $ f: {\Bbb R}^d \rightarrow [0,\infty) $ be a continuous function 
of compact support.
We will only consider the case where $L$ is so large that $\Lambda_L $
contains supp$\,f$.
We regard $f$ as a function on $\Lambda_L$ naturally.

Let  
\begin{equation}
       \tilde G_L = G_L^{1/2}e^{-f}G_L^{1/2},
\label{tildeg}
\end{equation}
where
$e^{-f}$ represents the operator of multiplication by the function $e^{-f}$.

Suppose there are $N$ identical particles which obey Bose-Einstein
statistics in $\Lambda_L$ under the periodic boundary condition
at inverse temperature $\beta$.
The basic postulates of quantum mechanics and of statistical mechanics of 
canonical ensembles yield 
\begin{equation}
      p^B_{L, N}(x_1, \cdots, x_N)
      = \frac{1}{Z_BN!}{\rm per}\,\{G(x_i, x_j)\}_{i,j=1}^N
\label{d+1}
\end{equation}
as the probability density distribution of the positions of $N$ particles of the system,
where $Z_B$ is the normalization constant and per represents the permanent of matrices.
Here, we have set $\hbar^2/2m = 1$.
We define the random point field ( the probability measure on $Q(\R^d)$ ) $ \mu_{L, N}^B $
induced by the map  
$ \Lambda_L^N \ni ( x_1, \cdots, x_N) \mapsto 
             \sum_{j=1}^N \delta_{x_j} \in Q(\R^d) $
from the probability measure on $\Lambda_L^N$ 
which has the density (\ref{d+1}).
By $ {\rm E}^B_{L,N} $, we denote the expectation with respect to $ \mu_{L, N}^B $. 
The Laplace transform of the point process is given by
\begin{eqnarray}
     {\rm E}^B_{L,N}\big[e^{-<f,\xi>}\big]
     &=& \frac{\int_{\Lambda^N}\exp(-\sum_{j=1}^Nf(x_j)){\rm per}\,\{G_L(x_i,
x_j)\}_{i,j=1}^N
     \,dx_1\cdots dx_N}
     {\int_{\Lambda^N}{\rm per}\,\{G_L(x_i, x_j)\}_{i,j=1}^N\,dx_1\cdots dx_N}
\notag \\
      &=& \frac{\int_{\Lambda^N}{\rm per}\,\{\tilde G_L(x_i, x_j)\}_{i,j=1}^N\,dx_1\cdots
dx_N}
     {\int_{\Lambda^N}{\rm per}\,\{G_L(x_i, x_j)\}_{i,j=1}^N\,dx_1\cdots dx_N}.
\label{bgfl}
\end{eqnarray}

Let us consider the thermodynamic limit, where $N$ and the volume 
of the box $\Lambda_L$ tend to infinity in such a way that the densities tend
 to a positive finite value $\rho $:
\begin{equation}
       L,\; N \rightarrow \infty, \quad N/L^d \to \rho > 0.
\label{tdl}
\end{equation}
In this paper, we concentrate on the high density region
\begin{equation}
      \rho > \rho_c = \int_{\R^d} \frac{dp}{(2\pi)^d}\frac{e^{-\beta |p|^2}}
              {1 - e^{-\beta |p|^2}}
\label{hd}
\end{equation}
where the Bose-Einstein condensation takes place.

\begin{thm}
\quad{\rm (i)} The operator $K=G(1-G)^{-1}$ is a non-negative unbounded 
self-adjoint operator in $L^2(\R^d)$ and satisfies Condition K in \S 2.
Moreover, $K_f$ defined by (\ref{K_f}) is a trace class operator.

{\rm (ii)} The finite point fields defined above converge weakly to the
random point field whose Laplace transform is given by
\begin{equation}
      {\rm E}_{\rho}^B\big[e^{-<f, \xi>}\big]
       = \frac{\exp\big(-(\rho-\rho_c)(\sqrt{1-e^{-f}}, [1+K_f]^{-1}\sqrt{1-e^{-f}})\big)}
              {\Det [1 + K_f]}
\label{BEC}
\end{equation}
in the thermodynamic limit (\ref{tdl}--\ref{hd}).
\label{TA}
\end{thm}
{\sl Remark:} Thus the resulting point field of the theorem is a convolution 
of a point field which is an example of those discussed in \S 2 and a boson 
process.
On the formulation of boson processes, we refer to \cite{ST}, where the operator
 $K$ is assumed to be bounded,  however the proof given there is also valid for 
the present case. 

\bigskip

Let us begin the proof with the following lemma, where we use the notation
\[
            \Box_k^{(L)} \; = \; \frac{2\pi}{L}\Big(k + 
          \Big( -\frac{1}{2}, \frac{1}{2}\Big]^d\Big)
         \qquad \mbox{for} \quad k \in \Z^d. 
\]
\begin{lem}
For $z \in [0, 1], \; \nu=1, 2 $ and $L \in [1,\infty)$, let us define
functions $a_{\nu}( \,\cdot \,;z), a_{\nu}^{(L)}(\,\cdot\,;z)$ on $\R^d$ by
\[
    a_{\nu}(p; z) = \frac{ze^{-\beta|p|^2}}{(1 - ze^{-\beta|p|^2})^{\nu}}
\]
and
\[
     a_{\nu}^{(L)}(p; z) =
     \begin{cases}
     0 & \mbox{ if } \quad p\in \Box_0^{(L)}
      \\
     a_{\nu}(2\pi k/L; z) & \mbox{ if } \quad
     p\in \Box_k^{(L)}  \quad \mbox{for} \quad k \in \Z^d -\{0\}.
     \end{cases}
\]
Then
\[
    0 \leqslant a_1^{(L)}(p;z)\leqslant a_1(2p/(2+\sqrt d);1) \in L^1(\R^d)
\]
and the bounds for large $L$
\[
  \frac{L^d}{(2\pi)^d} \int_{\R^d}a_{2}^{(L)}(p ;z)\,dp 
    \leqslant \ell(L) \equiv
   \begin{cases}c_d(L/\sqrt {\beta})^d & \mbox{ if } \quad d>4  \\
       \tilde c_4(L/\sqrt {\beta})^4\log(\tilde cL/\sqrt {\beta}) & 
      \mbox{ if } \quad d=4  \\
       c_d(L/\sqrt {\beta})^4 & \mbox{ if } \quad d<4 
       \end{cases}
\]
hold, where $c_d, \tilde c_4$ and $\tilde c$ are positive constants.
\label{A}
\end{lem}
\noindent{\sl Proof:}  
Since $a_{\nu}$ is monotone increasing in $z$ and monotone decreasing as 
a function of $|p|$, we have
\[
        a_{\nu}^{(L)}(p; z) \leqslant \sup_{L\geqslant 1}a_{\nu}^{(L)}(p; 1) \leqslant
        \sup\{ \, a_{\nu}(q; 1) \, | \, q\in \R^d, L\geqslant 1, |q|\geqslant \frac{2\pi}{L}, 
           \; |q - p|\leqslant (2\pi/L)(\sqrt d/2) \,\}
\]
\[
     \leqslant \sup\{ \, a_{\nu}(q; 1)  \,|\, q\in \R^d, L\geqslant 1, |q|\geqslant \frac{2\pi}{L}, 
           \; |p| - \pi\sqrt d/L \leqslant |q | \, \}.
\]
In the case of $ |p| \geqslant (2 + \sqrt d)\pi $, the last supremum is attained
at $ L=1, |q| = |p| - \pi\sqrt d $ then $ |q| \geqslant |2p|/(2+\sqrt d) $ holds.
On the other hand, if $ |p| < (2+ \sqrt d)\pi $, the supremum is attained
at $ L = (2+ \sqrt d)\pi/|p|, \; |q| = 2\pi/L$ and then 
$ |q| = |2p|/(2+\sqrt d) $ holds.
For both cases, we get the bound $  a_{\nu}^{(L)}(p;z)\leqslant a_{\nu}(2p/(2+\sqrt d);1)           $.
Since $d > 2$, we get $ a_1(2p/(2+\sqrt d); 1) \in L^1(\R^d)$.

Integrating the angular variables, we have
\[
        \frac{L^d}{(2\pi)^d} \int_{\R^d}a_2^{(L)}(p ;z)\,dp 
       \leqslant 
         \frac{L^d}{(2\pi)^d} \int_{|p|\geqslant\pi/L} a_2(2p/(2+\sqrt d) ;1) dp
\]
\[
       = \bigg(\frac{L}{2\pi\sqrt{\beta'}}\bigg)^d S_d 
 \int_{\pi\sqrt{\beta'}/L}^{\infty}\frac{q^{d-1}e^{-q^2}}{(1-e^{-q^2})^2}dq
        = \bigg(\frac{L}{2\pi\sqrt{\beta'}}\bigg)^d S_d {\cal I}_{d},
\]
where $\beta' = 4\beta/(2 + \sqrt d)^2$.
Since 
$ {\cal I}_{d} \leqslant \int_0^{\infty}[q^{d-1}e^{-q^2}/(1-e^{-q^2})^2]\,dq
  < \infty $  for $ d > 4$;
$ {\cal I}_{d} \leqslant \int_{\pi\sqrt{\beta'}/L}^{\infty}[q^{d-1}/q^4]\,dq
     = (4-d)^{-1}(L/\pi\sqrt{\beta'})^{4-d}$ for $ d < 4$
and
\begin{equation}
     {\cal I}_4 \leqslant \int_1^{\infty}\frac{q^{3-1}e^{-q^2}}{(1-e^{-q^2})^2}\,dq
              + \int_{\pi\sqrt{\beta'}/L}^{1}\frac{q^3}{q^4}\,dq
            = const. + \log \frac{L}{\pi\sqrt{\beta'}},
\end{equation}
we get the bounds for $\pi\sqrt{\beta} \leqslant L$. \hfill $\square$

\bigskip

\noindent ({\sl Proof  of Theorem \ref{TA}(i)}) It is obvious that $K=G(1-G)^{-1}$ is a 
unbounded non-negative self-adjoint operator satisfying $G=K(1+K)^{-1}$.
In fact, $K$ is explicitly given by the Fourier transformation:
\[
   K\phi = {\cal F}^{-1}(a_1(\,\cdot\,;1){\cal F}\phi)
\]
for 
\[
    \phi \in {\rm Dom\,}K = \{ \, \psi \in L^2(\R^d) 
          \, | \, a_1(\,\cdot\,;1){\cal F}\psi \in L^2(\R^d) \,\}.
\]
Condition K(ii) for $G$ are also obvious.

Let us show the locally boundedness of $K$.
For bounded measurable $\Lambda \subset \R^d$,
\[
    ||\sqrt K \chi_{\Lambda}\phi||^2_2  = ||\sqrt {a_1(\,\cdot\,;1)}
     {\cal F}(\chi_{\Lambda}\phi)||^2_2 \leqslant ||\sqrt{a_1(\,\cdot\,;1)}||_2^2 
        ||{\cal F}(\chi_{\Lambda}\phi)||_{\infty}^2
\]
\[
     \leqslant ||a_1(\,\cdot\,;1)||_1||\chi_{\Lambda}\phi||_1^2
       \leqslant (2\pi)^d \rho_c ||\chi_{\Lambda}||_2^2||\phi||_2^2.
\]
Thus $K^{1/2}\chi_{\Lambda}$ is bounded.
$K(x,y)$ in (\ref{Kxy}) is given by
\[
     K(x,y) =  \sum_{n=1}^{\infty}G^n(x,y) = \sum_{n=1}^{\infty}\int\frac{dp}{(2\pi)^d}
e^{-n\beta |p|^2+ip\cdot (x-y)}
\]
\[
   = \int\frac{dp}{(2\pi)^d}a_1(p;1)e^{ip\cdot (x-y)},
\]
where we have used the dominated convergence theorem.
From $a_1 \in L^1(\R^d)$, $K(x,y)$ is continuous. 
The remark after Proposition \ref{GB} and the continuity of 
$ f $ yield that the kernel of $K_f$ is continuous.
Hence $K_f$ is a trace class operator, 
because $ ||K_f||_T = \int K_f(x, x)\,dx = \rho_c||1-e^{-f}||_1 < \infty$. 
\hfill$\Box$

\medskip

The rest of this section is devoted to the proof of the second part of the theorem.
It is obvious from non-negativity of $f$ that
$ 0 \leqslant \tilde G_L \leqslant G_L $.   
Let us denote all the eigenvalues of $\tilde G_L$ in decreasing order
\[
        \tilde g_0(L) \geqslant \tilde g_1(L) \geqslant \cdots \geqslant
        \tilde g_j(L) \geqslant \cdots.
\]
Correspondingly, we relabel the eigenvalues
$ \, \{\, g_k^{(L)} \, \}_{k\in {\Bbb Z}^d} \,$ of $G_L$ as
\[
        g_0(L) =1 >  g_1(L) \geqslant \cdots \geqslant
         g_j(L) \geqslant \cdots.
\]
By the min-max principle, we have
\[
  g_j(L) \geqslant \tilde g_j(L) \qquad ( j = 0, 1, 2, \cdots ).
\]
Note that $\varphi_0^{(L)}$ has eigenvalue $g_0(L)=g_0^{(L)}=1$.

Put
\[
      D_L = G_L - \tilde G_L = G_L^{1/2}(1 - e^{-f})G_L^{1/2}, \quad
      W_L = G_L^{1/2}\sqrt{1-e^{-f}},
\]
then $\, D_L = W_LW_L^* $.
Note also that
\begin{equation}
  \frac{L^d}{(2\pi)^d} \int_{\R^d}a_{\nu}^{(L)}(p ;z)\,dp 
   = \sum_{k\in \Z^d-\{0\}}\frac{zg_k^{(L)}}{(1- zg_k^{(L)})^{\nu}}
  = ||zQ_0G_LQ_0
       (1 - zQ_0G_LQ_0)^{-\nu}||_T.
\label{a||||_T}
\end{equation}
Here $P_0$ is the orthogonal projection on $ L^2(\Lambda_L) $ to its one 
dimensional subspace ${\Bbb C}\varphi_0^{(L)}$ and $ Q_0 = 1 - P_0$.  

Now we have; 

\begin{lem}
\qquad {\rm (i)} \qquad  $ ||Q_0G_LQ_0(1 - Q_0G_LQ_0)^{-2}||_T 
              = \sum_{k\ne 0}g_k(L)/(1-g_k(L))^2  \leqslant \ell(L) $
\begin{eqnarray*}
\mbox{In the} & \mbox{limit}  L \to \infty, \mbox{the following 
convergences hold.} \hspace*{6cm}
\\
{\rm (ii)} &      L^{-d}\Tr G_L  \longrightarrow \int_{\R^d}e^{-\beta|p|^2}
                 dp/(2\pi)^d
               = \sqrt{4\pi\beta}^{-d},
   \quad    ||D_L||_T \longrightarrow ||1-e^{-f}||_1/
           \sqrt{4\pi\beta}^d
\\
{\rm (iii)} & L^{-d}||Q_0G_LQ_0(1 - Q_0G_LQ_0)^{-1}||_T 
             =  L^{-d}\sum_{k\ne 0}g_k(L)/(1-g_k(L)) 
 \longrightarrow   \rho_c < \infty
\\
{\rm (iv)} & \mbox{ If \; $\{z_L\} \subset (0,1)$  \; and \; $z_L \to 1$, \; then }
          \hspace*{6cm}
     \\
       & \sup_{x,y\in \Lambda} | [z_LQ_0G_LQ_0(1-z_LQ_0G_LQ_0)^{-1}](x,y) -  K(x,y) | \to 0
                     \\
        & \mbox{for any fixed bounded measurable set } \Lambda \subset \R^d.
\hspace*{5cm}
\end{eqnarray*}
\label{B}
\end{lem}

\medskip

{\sl Proof : } (i)--(iii) are immediate consequences of above remarks, Lemma \ref{A} 
and the dominated convergence theorem.

For (iv), put $e(p;x) = e^{ip\cdot x}$ and 
\[
      e^{(L)}(p; x) = e(2\pi k/L; x) \quad\mbox{if} \quad  
     p\in \Box_k^{(L)}
     \quad \mbox{for} \quad k \in \Z^d.
\]  
Then Lemma \ref{A} and the dominated convergence theorem also yields
\[
         | [z_LQ_0G_LQ_0(1-z_LQ_0G_LQ_0)^{-1}](x,y)  -  K(x,y) | 
         \leqslant \int\frac{dp}{(2\pi)^d}|e^{(L)}(p; x-y)a_1^{(L)}(p; z_L) -e(p; x-y)a_1(p; 1)|
\]
\begin{equation}
      \leqslant \int\frac{dp}{(2\pi)^d}\big(|a_1^{(L)}(p; z_L) - a_1(p; 1)| + 
      |e^{(L)}(p; x-y) - e(p; x-y)|a_1(p; 1)\big) \to 0.
\tag*{$\Box$}
\end{equation}  
\medskip

In the followings, we use the notation $B_L = \hat O(L^{\alpha})$ which means
\[
      \exists c_1 \geqslant c_2 > 0:  c_1 L^{\alpha} \geqslant B_L \geqslant c_2 L^{\alpha}.
\]
\begin{lem}
\begin{eqnarray*}
{\rm (i)}& &   \mbox{For large }L, \quad g_0(L) - \tilde g_0(L)    \\
        & &\hskip2cm = L^{-d}(\sqrt{1-e^{-f}},[1 + W_L^*Q_0[1-Q_0G_LQ_0]^{-1}Q_0W_L]^{-1}
                \sqrt{1-e^{-f}})\,(1 + o(1))   \\
  & & \hskip2cm = (\varphi_0^{(L)}, (D_L - D_LQ_0[1-Q_0\tilde G_LQ_0]^{-1}Q_0D_L)\varphi_0^{(L)})
              \,(1 + o(1)).   \\
{\rm (ii)} & & ||1-e^{-f}||_1/L^d = (\varphi_0^{(L)}, 
            D_L \varphi_0^{(L)}) \geqslant (\varphi_0^{(L)}, 
      D_LQ_0[\tilde g_0(L) - Q_0\tilde G_LQ_0]^{-1}Q_0D_L\varphi_0^{(L)})
              \\
{\rm (iii)}& &  \hskip2cm L^d(g_0(L) - \tilde g_0(L))  \in 
     \Big[ \frac{||1-e^{-f}||_1 (1+o(1))}{1+\rho_c||1-e^{-f}||_1},
     ||1-e^{-f}||_1\Big]
              \\
{\rm (iv)}& & \mbox{ Let } \tilde\varphi_0^{(L)} \mbox{ be the normalized eigenfunction of }
\tilde G_L  \mbox{ for eigenvalue } \tilde g_0(L) \mbox{ such that } \\
& &(\tilde\varphi_0^{(L)}, \varphi_0^{(L)}) \geqslant 0.  \mbox{ Put }
\tilde\varphi_0^{(L)} = a\varphi_0^{(L)} + \varphi', 
        \varphi_0^{(L)} = a'\tilde\varphi_0^{(L)} + \tilde\varphi' \quad 
  (\,(\varphi_0^{(L)}, \varphi') = 0, \quad 
\\
 & &   (\tilde\varphi^{(L)}_0, \tilde\varphi') = 0 \,).
 \mbox{ Then }
 a=a' \mbox{ and } ||\varphi'||^2 = ||\tilde\varphi'||^2 = 1-a^2
   = O(L^{-2d}\ell(L)) \mbox{ hold.}
\end{eqnarray*}
\label{C}
\end {lem}
{\sl Proof:} 
Here, we suppress the index $L$ in 
$ g_j(L), \tilde g_j(L), \varphi_0^{(L)} $ and so on.  
First notice that 
    $ (\varphi_0, D_L \varphi_0) = ||1-e^{-f}||_1/L^d $.
From the min-max principle, $d>2$ and the value of $g_1 = \exp(-\beta|2\pi/L|^2)$, we have
\begin{equation}
      g_0= 1 \geqslant \tilde g_0 \geqslant (\varphi_0, \tilde G_L \varphi_0)
      = 1 - (\varphi_0, D_L \varphi_0) = 1 - \hat O(L^{-d})
      > g_1 = 1 - \hat O(L^{-2}) \geqslant \tilde g_1 \geqslant \cdots
\label{V}
\end{equation}
for $L$ large enough.  
Hence the eigenspace of $\tilde G_L$ for its largest eigenvalue  $ \tilde g_0$ 
is one-dimensional.
Let $\tilde \varphi_0 $ be the normalized eigenfunction for $\tilde g_0$ and put 
     $\tilde\varphi_0 = a\varphi_0 + \varphi' \quad 
     (\,(\varphi_0, \varphi') = 0 \,)$, 
then $ \tilde G_L\tilde\varphi_0 = \tilde g_0\tilde\varphi_0 $ yields
\[
      a\tilde G_L\varphi_0 + \tilde G_L\varphi' = a\tilde g_0\varphi_0
      + \tilde g_0\varphi'.
\]
Applying $P_0$ and $Q_0$, we have
\begin{eqnarray*}
      ag_0 - a (\varphi_0, D_L\varphi_0) - (\varphi_0, D_L\varphi') 
      &=& a\,\tilde g_0      \\
      -aQ_0D_L\varphi_0 + Q_0 \tilde G_L\varphi' &=& \tilde g_0\,\varphi'.
\end{eqnarray*}
Because of $ Q_0\tilde G_LQ_0 \leqslant Q_0 G_LQ_0 \leqslant g_1 < \tilde g_0$,
$\tilde g_0 - Q_0\tilde G_LQ_0 $ is positive, invertible and 
\begin{eqnarray}
       \varphi' & = & -a[\tilde g_0 -Q_0\tilde G_LQ_0]^{-1}Q_0D_L\varphi_0, 
 \label{varphi'}\\
      g_0 - \tilde g_0 
            & = & (\varphi_0, (D_L - D_LQ_0[\tilde g_0-Q_0\tilde G_LQ_0]^{-1}
             Q_0D_L)\varphi_0)
  \notag\\
           & = & (W_L^*\varphi_0,(1 -  W_L^*Q_0
           [\tilde g_0-Q_0\tilde G_LQ_0]^{-1}Q_0W_L)W_L^*\varphi_0) .  
\label{i}
\end{eqnarray}
  
For brevity, we put
\[
X' = W_L^*Q_0[\tilde g_0-Q_0G_LQ_0]^{-1}Q_0W_L, \quad 
X  =  W_L^*Q_0[1 - Q_0G_LQ_0]^{-1}Q_0W_L
\]
and
\[
  \tilde X = W_L^*Q_0[\tilde g_0-Q_0\tilde G_LQ_0]^{-1}Q_0W_L.
\]
Then we have
\[
  \tilde X - X' = - \tilde X X',
\]
and hence
\begin{equation}
    \tilde X =  X'(1+X')^{-1} \quad \mbox{and} \quad 1-\tilde X = (1+X')^{-1}.
\label{XX'}
\end{equation}
Together with $ W_L^*\varphi_0 = \sqrt{1-e^{-f}}L^{-d/2} $, we have
\begin{equation}
      g_0 - \tilde g_0    
             =  L^{-d}(\sqrt{1-e^{-f}},(1 + X')^{-1} \sqrt{1-e^{-f}})
\label{f}
\end{equation}
from (\ref{i}).  
Now, we want to replace $X'$ by $X$ in the right hand side.
From 
$
     1 - \tilde g_0 =  O(L^{-d}), \; \tilde g_0 - g_1 = \hat O(L^{-2}), 
   \;  \sum_{k\ne 0}g_k/(1-g_k)^2 \leqslant \ell(L)
$
 ((\ref{V}), Lemma \ref{B}(i)), it follows that
\begin{eqnarray*}
    || X' - X || & = & (1-\tilde g_0)||W_L^*Q_0[\tilde g_0-Q_0G_LQ_0]^{-1}
       [1-Q_0G_LQ_0]^{-1}Q_0W_L||  
\\
       & = & (1-\tilde g_0)\sup_{||\phi ||_2=1} 
       (\phi, W_L^*Q_0[\tilde g_0-Q_0G_LQ_0]^{-1}[1-Q_0G_LQ_0]^{-1}Q_0W_L\phi)   
\end{eqnarray*}
\begin{eqnarray}
      & \leqslant & (1-\tilde g_0) \sup_{||\phi ||_2=1} \sum_{k \ne 0} 
       |(\varphi_k,\sqrt{1-e^{-f}}\phi)|^2 
       \frac{g_k}{(\tilde g_0 -g_k)(1-g_k)}
\notag\\
      & \leqslant & (1-\tilde g_0) \sup_{||\phi ||_2=1}  
      \frac{||\sqrt{1-e^{-f}}\phi||_1^2}{L^d} 
      \frac{1-g_1}{\tilde g_0 -g_1} \sum_{k \ne 0}\frac{g_k}{(1-g_k)^2}
\notag\\
      & = & ||1-e^{-f}||_1  O(L^{-2d}\ell(L)) = o(1).
\label{X'-X}
\end{eqnarray}
Together with the similar estimate  $ || X || \leqslant \rho_c||1-e^{-f}||_1(1+o(1))$,
we have $ || X' || \leqslant \rho_c||1-e^{-f}||_1(1+o(1))$.
Thus (\ref{f}) yields
\[
   L^d(g_0 - \tilde g_0) = (\sqrt{1-e^{-f}}, ( 1+ X' )^{-1} \sqrt{1-e^{-f}}) 
   \geqslant \frac{||1-e^{-f}||_1}{1+\rho_c||1-e^{-f}||_1}(1+o(1)),
\]
which is the lower bound of (iii).  The upper bound of (iii) is obvious.
From
\[
     |(\sqrt{1-e^{-f}}, ( 1+ X' )^{-1} \sqrt{1-e^{-f}})
     - (\sqrt{1-e^{-f}}, ( 1+ X )^{-1} \sqrt{1-e^{-f}})|
\]
\[
     \leqslant ||\sqrt{1-e^{-f}}||_2^2 ||(1 + X)^{-1}||\,||(1 + X')^{-1}||\,||X-X'||
     = o(1),
\]
we get the first equality of (i).
Replacing $X'$ by $X$ in (\ref{f}) and tracing the argument back to (\ref{i}),
we get the second one of (i).
(ii) is an immediate consequence of $g_0 \geqslant \tilde g_0$ and (\ref{i}).  

(iv) Clearly, $ a= (\tilde\varphi_0, \varphi_0) = a'$. 
As for (\ref{XX'}), we have
\begin{equation}
 (\tilde g_0 - Q_0\tilde G_LQ_0)^{-1}Q_0W_L 
            =  (\tilde g_0 - Q_0 G_LQ_0)^{-1}Q_0W_L(1+X')^{-1}.
\label{XX}
\end{equation}
This and estimates similar to (\ref{X'-X}) derive the bound 
\begin{eqnarray*}
       ||\varphi' ||^2 &=& a^2 (\varphi_0, D_LQ_0[\tilde g_0 - 
       Q_0\tilde G_LQ_0]^{-2}Q_0D_L\varphi_0 )      \\
       &\leqslant& a^2 ||W_L^*\varphi_0||_2^2 ||W_L^*Q_0[\tilde g_0 - Q_0\tilde G_LQ_0]^{-2}
       Q_0W_L||      \\
       &=& a^2 ||W_L^*\varphi_0||_2^2 ||(1+ X')^{-1}W_L^*Q_0[\tilde g_0 - Q_0 G_LQ_0]^{-2}
       Q_0W_L(1 + X')^{-1}||   \\
      &=& a^2O(L^{-2d}\ell(L))
\end{eqnarray*}
from (\ref{varphi'}).  
Now the bound for ${1-a^2}$ is obvious.
\hfill$\Box$

As in I, we use the generalized Vere-Jones' formula \cite{V, ST} in the form
\[
     \frac{1}{N!}\int 
                  \per(J(x_{i} ,x_{j}))_{i,j=1}^{N}
                  \lambda^{\otimes N}
                  (dx_{1}\cdots dx_{N})
      = \oint _{S_r(0)}\frac{dz}{2\pi iz^{N+1}}\Det(1- z J)^{-1},
\]
where $r>0$ satisfies $||r J||<1$.
$S_r(\zeta)$ denotes the integration contour defined
by the map  $ \theta \mapsto \zeta + r\exp(i\theta) $, 
where $\theta$ ranges from $-\pi$ to $\pi$, $ r>0 $ and $ \zeta \in \C$.
Then we get 
\begin{equation}
     {\rm E}_{L, N}^B\big[e^{-<f, \xi>}\big] = \frac{z_0^N}{\tilde z_0^N}
     \frac{\Det[1-z_0G_L]}{\Det[1-\tilde z_0\tilde G_L]}
     \frac{\oint_{S_1(0)}\Det\big[1-\tilde z_0\tilde G_L
          (1-\tilde z_0\tilde G_L)^{-1}(\eta-1)\big]^{-1}d\eta/2\pi
i\eta^{N+1}}
          {\oint_{S_1(0)}\Det\big[1- z_0 G_L
          (1 - z_0G_L)^{-1}(\eta-1)\big]^{-1}d\eta/2\pi i\eta^{N+1}}.
\label{D/Db}
\end{equation}
The positive real numbers $z_0=z_0(L,N)$ and
 $\tilde z_0 = \tilde z_0(L,N)$ are chosen as the solutions of the equations
\begin{equation}
   \Tr_{\H}\big[z_0G_L(1-z_0G_L)^{-1}\big] =
   \Tr_{\H}\big[\tilde z_0\tilde G_L(1-\tilde z_0\tilde G_L)^{-1}\big] = N.
\label{nb}
\end{equation}
In fact, the following lemma holds.
\begin{lem}
{\rm (i)} \quad $z_0 = z_0(L,N) \in (0, 1)$ is uniquely determined by the
equation
\begin{equation}
   \Tr\big[ z_0 G_L(1 - z_0 G_L)^{-1}\big] = N.
\label{H2}
\end{equation}
{\rm (ii)} \quad $\tilde z_0 = \tilde z_0(L,N) \in ( 0, \tilde
g_0^{-1}(L)\,)$ is
uniquely determined by the equation
\begin{equation}
   \Tr\big[\tilde z_0\tilde G_L(1 - \tilde z_0\tilde G_L)^{-1}\big] =
N.
\label{tildeH}
\end{equation}
{\rm (iii)} \quad $ 0\leqslant \tilde z_0 - z_0 = O(L^{-d})$

\noindent{\rm (iv)} \quad $1-z_0 = (1+o(1))L^{-d}(\rho-\rho_c)^{-1}$
\label{D}
\end{lem}
{\sl Proof : } Let $H(z_0)$ and $\tilde H(\tilde z_0)$ be the left-hand sides of 
(\ref{H2}) and  (\ref{tildeH}) , respectively.  
Since $H$ is monotone increasing continuous function, $H(0)=0$ and
$H(1-0)=\infty$, (i) follows.
(ii) is similar. 
The first inequality of (iii) is a consequence of $ H(z) \geqslant \tilde H(z)$.
Before showing the second inequality, let us make the following remark on the 
thermodynamic limit (\ref{tdl}).

\medskip

\noindent (a) {\sl If and only if $ \rho < \rho_c $,
$ \{ z_0( L, N) \} $ converges to  $ z = z_* \in ( 0, 1)$, the unique solution of}
\[
  \rho = \int \frac{dp}{(2\pi)^d}a_1(p;z).
\]

\noindent (b) {\sl If and only if $ \rho > \rho_c $,
$ L^d(1-z_0) \longrightarrow 1/(\rho -\rho_c)$, hence} $\lim z_0
=1$.

\noindent (c) {\sl If and only if $ \rho = \rho_c $, $\lim z_0 =1$ and}
             $ L^d(1-z_0) \longrightarrow +\infty.$
\medskip

To show (a -- c), note that
\[
    \frac{z_0( L, N)}{ L^d(1 - z_0( L, N))} +
    \int_{\R^d} \frac{dp}{(2\pi)^d } a_1^{(L)}(p; z_0(L, N))
\]
\begin{equation}
   = \Tr[z_0(L, N)G_L( 1 - z_0(L, N)G_L)^{-1}]/L^d
  = N/L^d \to \rho .
\label{btdl}
\end{equation}
We have that
\[
      \int_{\R^d} \frac{dp}{(2\pi)^d } a_1^{(L)}(p; z_0)
       \to \int_{\R^d} \frac{dp}{(2\pi)^d } a_1(p; z)
\]
for $\lim z_0 = z \in [0,1]$ by the dominated convergence theorem,
and that the limit is a strictly increasing function of $z$.
( See Lemma \ref{A}. )
If $\lim z_0 = z_* \in [ 0, 1)$, the limit of (\ref{btdl}) tends to
\[
        \rho = \int_{\R^d} \frac{dp}{(2\pi)^d } a_1(p; z_*) < \rho_c.
\]
If $ \lim z_0 = 1$, then $ \rho = \rho_c + \lim z_0/L^d(1-z_0) \geqslant \rho_c.$
Now suppose $ \{ z_0(L,N) \} $ does not converge. 
Then by taking converging subsequences having different limits,
we deduce a contradiction to (\ref{btdl}).
Thus we get the classification (a -- c) and (iv).

Now we have the second part of (iii) using  Lemma \ref{C}(iii),
\begin{equation}
       z_0 = 1 - \hat O(L^{-d}) \leqslant \tilde z_0 < \tilde g_0^{-1} = 1 +
\hat O(L^{-d}).
\tag*{$\Box$}
\end{equation}
\medskip

\medskip

In order to understand the subsequent arguments, it is helpful to keep 
the followings in mind:
\[
    g_0 = 1 , \quad g_1 = 1-\hat O(L^{-2}) \geqslant Q_0G_LQ_0 \geqslant Q_0\tilde G_LQ_0
     \qquad ( {\rm see \; (\ref{V})} )
\]
\[
    \tilde g_0 = 1 - \hat O(L^{-d})  \qquad ( {\rm Lemma \; \ref{C}(iii)} )
\]
\[
    z_0 = 1 - \hat O(L^{-d})  \quad \tilde z_0 = z_0 + O(L^{-d})
      \qquad ( {\rm Lemma \; \ref{D}(iii, iv)} )
\]
\[
       (\varphi_k^{(L)}, D_L\varphi_k^{(L)}) = g_k^{(L)}||1-e^{-f}||_1/L^d
\]
\begin{lem}
  \begin{eqnarray*}
     {\rm (i)}&           & P_0[1-\tilde z_0 \tilde G_L]^{-1}P_0
           = \Big(\frac{1}{1-\tilde z_0 \tilde g_0} + O(L^{-d}\ell(L))\Big)P_0,
              \\
     {\rm (ii)}&          & ||Q_0[1-\tilde z_0 \tilde G_L]^{-1}|| 
                    = ||[1-\tilde z_0 \tilde G_L]^{-1}Q_0|| = O(\sqrt{\ell(L)}),
               \\
               &           & ||Q_0[1- z_0 \tilde G_L]^{-1}|| 
                      = ||[1- z_0 \tilde G_L]^{-1}Q_0|| = O(\sqrt{\ell(L)}),
               \\
     {\rm (iii)}&         & \Tr (Q_0[1 - z_0 G_L]^{-1}D_L
                     [1 - z_0 G_L]^{-1}Q_0) = O(L^{-d}\ell(L)).
\end{eqnarray*}
\label{E}
\end{lem}

\noindent{\sl Proof: }
(i) \quad By lemma \ref{C}(iv), we have
\[
   |(\varphi_0,(1-\tilde z_0\tilde G_L)^{-1}\varphi_0)
          - (1-\tilde z_0\tilde g_0)^{-1}|
        = |(a\tilde\varphi_0 + \tilde\varphi',(1-\tilde z_0\tilde G_L)^{-1}
          (a\tilde\varphi_0 + \tilde\varphi')) - (1-\tilde z_0\tilde
g_0)^{-1}|
\]
\[
          \leqslant \frac{1-a^2}{1-\tilde z_0\tilde g_0} +
           |(\tilde\varphi', (1-\tilde z_0\tilde G_L)^{-1}\tilde\varphi')|
          \leqslant ((1-\tilde z_0\tilde g_0)^{-1}+(1-\tilde z_0\tilde g_1)^{-1})
          O(L^{-2d}\ell(L)) = O(L^{-d}\ell(L)),
\]
where we have used
\[
       \frac{1}{1-\tilde z_0\tilde g_0} + \frac{1}{1-\tilde z_0\tilde g_1}
     \leqslant 2+ \Tr[\tilde z_0\tilde G_L(1-\tilde z_0\tilde G_L)^{-1}] = 2+N = O(L^d),
\]
in the last step.

\noindent (ii) \quad Note that $ Q_0\tilde \varphi_0 =\varphi' $ in the
notation of lemma \ref{C}(iv). Then we get
\[
         ||Q_0 (1-\tilde z_0\tilde G_L)^{-1}|| 
      \leqslant  \frac{||\varphi'||}{1- \tilde z_0\tilde g_0}
          + \frac{1}{1-\tilde z_0\tilde g_1} = O(L^d\sqrt{L^{-2d}\ell(L)}) + O(L^2).
\]
The second bound is obtained similarly.

\noindent (iii) \quad From the above remark, the left-hand side equals
\[
          \frac{||1-e^{-f}||_1}{L^d}
           \sum_{k\ne 0}\frac{g_k}{(1-z_0g_k)^2},
\]
which yields the righthand side by Lemma \ref{B}(i).  \hfill$\square$

\medskip

\noindent We need a finer estimate than Lemma \ref{D}(iii).
\begin{lem}
  \begin{eqnarray*}
   {\rm (i)}& \hskip1cm  & \tilde z_0 -z_0 = ( 1-\tilde g_0)(1+o(1)),
                      \\
   {\rm (ii)}&          &  1-\tilde z_0g_0' = (1-\tilde z_0\tilde g_0)(1+o(1))
                    = (1-z_0)(1+o(1)) 
                     =\frac{1+o(1)}{L^d(\rho -\rho_c)},
                     \\
         \mbox{where} & g_0' = 1&- (\varphi_0^{(L)}, D_L \varphi_0^{(L)}) +
       \tilde z_0 (\varphi_0^{(L)},
      D_LQ_0[1 - \tilde z_0Q_0\tilde G_LQ_0]^{-1}Q_0D_L\varphi_0^{(L)}).
  \end{eqnarray*}
\label{F}
\end{lem}
{\sl Proof: } (i) \quad Let us begin with
\[
        0=N-N=\Tr[\tilde z_0\tilde G_L(1-\tilde z_0\tilde G_L)^{-1}
                   - z_0G_L(1-z_0G_L)^{-1}]
\]
\[
         = (\varphi_0, ((1-\tilde z_0\tilde G_L)^{-1}
                  -(1-z_0G_L)^{-1})\varphi_0)
         + \Tr[Q_0((1-\tilde z_0\tilde G_L)^{-1} - (1-z_0\tilde G_L)^{-1})Q_0]
\]
\[
           + \Tr[Q_0((1- z_0\tilde G_L)^{-1} -  (1-z_0G_L)^{-1})Q_0].
\]
The first term of the right hand side equals
\[
        (1-\tilde z_0\tilde g_0)^{-1} -(1-z_0g_0)^{-1} +O(L^{-d}\ell(L))
        = \frac{(\tilde z_0 - z_0)\tilde g_0 - z_0( g_0 - \tilde g_0)}
          {(1-\tilde z_0\tilde g_0)(1-z_0g_0)} +O(L^{-d}\ell(L))
\]
by Lemma \ref{E}(i).  On the other hand, the second term has the bound
\[
  (\tilde z_0 -z_0)|\Tr[Q_0(1-\tilde z_0\tilde G_L)^{-1}\tilde G_L(1-z_0\tilde G_L)^{-1}Q_0]| 
\]
\[
    \leqslant \frac{\tilde z_0 -z_0}{\tilde z_0} 
    ||\tilde z_0\tilde G_L(1-\tilde z_0\tilde G_L)^{-1}||_T
        ||(1- z_0\tilde G_L)^{-1}Q_0|| = O(L^{-d}L^d\sqrt{\ell(L)})=o(L^d)
\]
by Lemma \ref{D}(iii, ii) and Lemma \ref{E}(ii).  
The third term can be estimated as
\[
    |\Tr[Q_0((1- z_0\tilde G_L )^{-1}-(1-z_0G_L)^{-1})Q_0]|
    = z_0 |\Tr[Q_0(1- z_0\tilde G_L )^{-1}W_LW_L^*(1-z_0G_L)^{-1}Q_0]|
\]
\[
     = z_0|\Tr[Q_0(1- z_0 G_L)^{-1}W_L(1+ z_0W_L^*(1-z_0G_L)^{-1}W_L)^{-1}
        W_L^*(1-z_0G_L)^{-1}Q_0]|
\]
\[     
          \leqslant z_0 ||Q_0(1- z_0G_L)^{-1}W_LW_L^*(1-z_0G_L)^{-1}Q_0]||_T  
         = O(L^{-d}\ell(L)) = o(L^d),
\]
where we have used a equality similar to (\ref{XX}) and Lemma \ref{E}(iii).
Thus we have
\[
    \frac{z_0(g_0-\tilde g_0)-(\tilde z_0 -z_0)\tilde g_0}
    {(1-\tilde z_0\tilde g_0)(1-z_0g_0)} = o(L^d).
\]
On the other hand, $ (1-\tilde z_0\tilde g_0)(1-z_0g_0) =O(L^{-2d})$ holds.
Thus we have
\[
   z_0(g_0-\tilde g_0)-(\tilde z_0 -z_0)\tilde g_0 =o(L^{-d}).
\]
Note that $ g_0-\tilde g_0 $ is exactly of order $L^{-d}$ by
 Lemma \ref{C}(iii), we get the desired estimate.

\noindent (ii) \quad From (\ref{i}), we have
\[
    |\tilde g_0 -g_0'|=|(\varphi_0, D_LQ_0[(\tilde g_0-Q_0\tilde G_LQ_0)^{-1}
         -(\tilde z_0^{-1}-Q_0\tilde G_LQ_0)^{-1}]Q_0D_L\varphi_0)|
\]
\[
          =|(\varphi_0, D_LQ_0(\tilde g_0-Q_0\tilde G_LQ_0)^{-1/2}
         [(\tilde z_0^{-1}-\tilde g_0)(\tilde z_0^{-1}-Q_0\tilde
G_LQ_0)^{-1}]
             (\tilde g_0-Q_0\tilde G_LQ_0)^{-1/2}Q_0D_L\varphi_0)|
\]
\[
           \leqslant   |\tilde z_0^{-1}-\tilde g_0|
              ||(\tilde z_0^{-1}-Q_0\tilde G_LQ_0)^{-1}||
         (\varphi_0,D_LQ_0(\tilde g_0-Q_0\tilde G_LQ_0)^{-1}Q_0D_L\varphi_0)
\]
\[
          \leqslant O(L^{-d})O(L^2)(\varphi_0,D_L\varphi_0)=O(L^{2-2d})=o(L^{-d}),
\]
where Lemma \ref{C}(ii) has been used in the last inequality.
Hence, we obtain $ 1-\tilde z_0g_0' = 1-\tilde z_0\tilde g_0 + o(L^{-d})$.
On the other hand, we have
\[
      1-\tilde z_0\tilde g_0 =1-z_0+[\tilde z_0(1-\tilde g_0)-(\tilde
z_0-z_0)]
\]
\[
         = \frac{1+o(1)}{L^d(\rho -\rho_c)}+ o(L^{-d}),
\]
thanks to Lemma \ref{D}(iv) and (i) above.
\hfill$\square$

\bigskip

Put
\[
    p_j^{(N)} = \frac{z_0(L,N)g_j(L)}{1 - z_0(L,N)g_j(L)},
    \qquad \tilde p_j^{(N)} = \frac{\tilde z_0(L,N)\tilde g_j(L)}
    {1 - \tilde z_0(L,N)\tilde g_j(L)},
\]
then we have
$\sum_{j=0}^{\infty}p_j^{(N)}=\sum_{j=0}^{\infty}\tilde p_j^{(N)}=N$ by
Lemma \ref{D}(i, ii),
\begin{equation}
p_0^{(N)}=\hat O(L^d), \quad \tilde p_0^{(N)}=\hat O(L^d), \quad
p_0^{(N)}/\tilde p_0^{(N)} =1+o(1)
\label{p/p}
\end{equation}
by Lemma \ref{F}(ii) and
\[
      p_1^{(N)}= \hat O(L^2) \geqslant p_2^{(N)} \geqslant \cdots,\qquad
     \tilde p_1^{(N)}= O(L^2) \geqslant \tilde p_2^{(N)} \geqslant \cdots.
\]
\begin{lem}
\begin{eqnarray*}
\oint_{S_1(0)}\frac{1}{\Det\big[1- z_0(L,N)G_L
          (1-z_0(L,N)G_L)^{-1}(\eta-1)\big]}
          \frac{d\eta}{2\pi i\eta^{N+1}} &=& \frac{1+o(1)}{e p_0^{(N)}}
  \\
\oint_{S_1(0)}\frac{1}{\Det\big[1-\tilde z_0(L,N)\tilde G_L
          (1-\tilde z_0(L,N)\tilde G_L)^{-1}(\eta-1)\big]}
          \frac{d\eta}{2\pi i\eta^{N+1}} &=& \frac{1+o(1)}{e\tilde p_0^{(N)}}
\end{eqnarray*}
\label{G}
\end{lem}
{\sl Proof : }
Set $R^{(N)}=\tilde R^{(N)}=L^{(d-2)/2}$.
Since $
\sum_{j=1}^{\infty}p^{(N)}_j(1+p^{(N)}_j)=\Tr[z_0Q_0G_LQ_0(1-z_0Q_0G_LQ_0)^{-2}]
 \leqslant \sum_{j=1}^{\infty}g_j/(1-g_j)^2 $,
we get
\[
    \frac{R^{(N)2}\sum_{j=1}^{\infty}p^{(N)}_j(1+p^{(N)}_j)}{p^{(N)2}_0} 
    \to 0
\]
by $p^{(N)}_0 =\hat O(L^d)$ and  Lemma \ref{B}(i).
Then Lemma A.2 yields
\[
   \mbox{the l.h.s. of the 1st eq.} = \oint_{S_1(0)}
     \frac{1}{\prod_{j=0}^{\infty}(1- p_j^{(N)}(\eta-1))}
   \frac{d\eta}{2\pi i\eta^{N+1}} = \frac{1+o(1)}{e p_0^{(N)}}.
\]

For the second equality, we notice that $\tilde p_j^{(N)} \leqslant (1+o(1))
p_j^{(N)}$
holds for all $j=1, 2, \cdots$, because of
$z_0, \tilde z_0 = 1 + O(L^{-d})$ and 
$\tilde g_j^{(N)} \leqslant g_j^{(N)} \leqslant 1-\hat O(L^{-2})$.
Together with (\ref{p/p}), we have
\[
   \frac{\tilde R^{(N)2}\sum_{j=1}^{\infty}\tilde p^{(N)}_j(1+\tilde p^{(N)}_j)}
   {\tilde p^{(N)2}_0}
  \leqslant (1+o(1)) \frac{R^{(N)2}\sum_{j=1}^{\infty}p^{(N)}_j(1+p^{(N)}_j)}{p^{(N)2}_0} 
  \to 0.
\]
Thus the second equality also follows from Lemma A.2.
\hfill$\square$

\bigskip

Now we have
\[
    {\rm E}_{L, N}^B\big[e^{-<f, \xi>}\big]
         = \frac{z_0^{N}}{\tilde z_0^{N}}
     \frac{\Det[1- z_0G_L]}{\Det[1-\tilde z_0\tilde G_L]}(1+o(1))
\]
from (\ref{D/Db}), (\ref{p/p}) and the above lemma.
Since $P_0, Q_0$ and $G_L$ commute, $ \Det[1-z_0G_L]=(1-z_0)\Det[1-z_0Q_0G_LQ_0]$.
We use the Feshbach formula to get
\[
    \Det[1-\tilde z_0\tilde G_L]
      = \Det
   \begin{pmatrix}P_0 -\tilde z_0P_0\tilde G_LP_0 & -\tilde z_0P_0\tilde
G_LQ_0 \\
            -\tilde z_0Q_0\tilde G_LP_0 & Q_0-\tilde z_0Q_0\tilde G_LQ_0
          \end{pmatrix}
\]
\[
      = \Det_{Q_0\H_L}[Q_0-\tilde z_0Q_0\tilde G_LQ_0 ]
\]
\[
       \times \Det_{P_0\H_L}[P_0-\tilde z_0P_0\tilde G_LP_0
     -\tilde z_0P_0\tilde G_LQ_0(Q_0-\tilde z_0Q_0\tilde G_LQ_0 )^{-1}
    \tilde z_0Q_0\tilde G_LP_0]
\]
\[
       = \Det [1-\tilde z_0Q_0\tilde G_LQ_0 ]
\]
\[
       \times \big(1-\tilde z_0[1- (\varphi_0^{(L)}, D_L \varphi_0^{(L)}) +
       \tilde z_0 (\varphi_0^{(L)}, D_LQ_0[1 - \tilde z_0Q_0\tilde G_LQ_0
      ]^{-1}Q_0D_L\varphi_0^{(L)})]\big)
\]
where  Det is the Fredholm determinant for operators on $\H_L$ and
$\Det_{Q_0\H_L}$ for operators on the subspace $Q_0\H_L$ etc.
Now from Lemma \ref{F}(ii) and Lemma \ref{D}(iii, iv), we get
\[
        {\rm E}_{L, N}^B\big[e^{-<f, \xi>}\big]
     =  \frac{z_0^{N}}{\tilde z_0^{N}}
      \frac{(1-z_0)\Det[1- z_0Q_0G_LQ_0]}
     {(1 - \tilde z_0 g_0')\Det[1-\tilde z_0Q_0\tilde G_LQ_0]}(1+o(1))
\]
\[
    = \frac{z_0^N}{\tilde z_0^N}
     \frac{\Det[1- z_0Q_0G_LQ_0]}
     {\Det[1-\tilde z_0Q_0\tilde G_LQ_0]}(1+o(1))
\]
\begin{equation}
    = \exp\big(-\frac{\tilde z_0-z_0}{z_0}N+o(1)\big)
     \frac{\Det[1- z_0Q_0G_LQ_0]}{\Det[1-Q_0G_LQ_0]}
      \frac{\Det[1- Q_0G_LQ_0]}{\Det[1-Q_0\tilde G_LQ_0]}
     \frac{\Det[1- Q_0\tilde G_LQ_0]}{\Det[1-\tilde z_0Q_0\tilde G_LQ_0]}.
\label{DDD}
\end{equation}

\medskip
\begin{lem}
\begin{eqnarray*}
{\rm (i)}& \hskip1cm &  \frac{\Det[1- z_0Q_0G_LQ_0]}{\Det[1-Q_0G_LQ_0]}
                      = \exp\big(\frac{1-z_0}{z_0}(N-p_0^{(N)}) +o(1)\big)
               \\
{\rm (ii)}&         &   \frac{\Det[1- \tilde z_0Q_0\tilde G_LQ_0]}
                       {\Det[1-Q_0\tilde G_LQ_0]}
          = \exp\big(\frac{1-\tilde z_0}{\tilde z_0}(N-\tilde p_0^{(N)})
+o(1)\big)
              \\
{\rm (iii)}&        & \frac{\Det[1- Q_0\tilde G_LQ_0]}{\Det[1-Q_0G_LQ_0]}
                = \Det[1+K_f](1+o(1))
\end{eqnarray*}
\label{H}\end{lem}
{\sl Proof : } Put $ h(z) = -\log\Det (1-zQ_0G_LQ_0)=-\sum_{j=1}^{\infty}
\log(1-zg_j)$, and we have
\[
        \log\frac{\Det[1- z_0Q_0G_LQ_0]}{\Det[1-Q_0G_LQ_0]}
        = h(1)-h(z_0)=h'(z_0)(1-z_0)+\frac{1}{2}h''(\bar z_0)(1-z_0)^2,
\]
where $\bar z_0 \in (z_0, 1)$.
Hence we get (i) by
\[
    h'(z_0)=\sum_{j=1}^{\infty}\frac{g_j}{1-z_0g_j}=\frac{N-p_0}{z_0}
\]
and
\[
   h''(\bar z_0)(1-z_0)^2=\sum_{j=1}^{\infty}\frac{g_j^2(1-z_0)^2}{(1-\bar
z_0g_j)^2}
   \leqslant \sum_{j=1}^{\infty}\frac{g_j(1-z_0)^2}{(1-g_j)^2}= O(L^{-2d}\ell(L)) = o(1),
\]
where Lemma \ref{B}(i) has been used.
Similar argument and
$ \sum_{j=1}^{\infty}\tilde p_j(1+\tilde p_j)
  \leqslant (1+o(1)) \sum_{j=1}^{\infty}p_j(1+p_j)  $
yield (ii).

\noindent(iii) Thanks to the product and cyclic properties of the Fredholm determinant, 
we have
\[
\frac{\Det[1-Q_0\tilde G_LQ_0]}{\Det[1-Q_0G_LQ_0]}
           = \Det [1 +Q_0(G_L-\tilde G_L)Q_0(1-Q_0G_LQ_0)^{-1}]
           = \Det [1 + W_L^*Q_0(1-Q_0G_LQ_0)^{-1}Q_0W_L]
\]
\[
      = \Det [1 + \sqrt{1-e^{-f}}Q_0G_LQ_0(1-Q_0G_LQ_0)^{-1}
           \sqrt{1-e^{-f}}].
\]
Note that $ L^2(\Lambda_L)$ can be identified with an closed subspace 
of $L^2(\R^d)$ naturally.
By this identification, we regard $ G_L$ and $ \sqrt{1-e^{-f}}$ as operators
 on $L^2(\R^d)$.
Now for (iii), it is enough to prove 
\[
      A_L = \sqrt{1-e^{-f}}Q_0G_LQ_0(1-Q_0G_LQ_0)^{-1}
           \sqrt{1-e^{-f}} \longrightarrow K_f  
\]
in the trace norm.
In the following,  we show $A_L \to K_f$ strongly and $||A_L||_T \to ||K_f||_T$. 
Then the Gr\"um's convergence theorem \cite{Si} yields the above.

For $\psi, \phi \in L^2(\R^d)$, we have
\[
   | (\psi,(A_L - K_f )\phi)| = \bigg|\int_{\R^d}dx\int_{\R^d}dy \,
    \overline{\psi(x)}\sqrt{1-e^{-f(x)}}
    \phi(y)\sqrt{1-e^{-f(y)}}
\]
\[
      \big([Q_0G_LQ_0(1-Q_0G_LQ_0)^{-1}](x,y) - K(x,y)\big)\bigg|
\]
\begin{equation}
    \leqslant (|\psi|,\sqrt{1-e^{-f}})
    (\sqrt{1-e^{-f}}, |\phi|)
   \sup_{x, y \in \supp f}|[Q_0G_LQ_0(1-Q_0G_LQ_0)^{-1}](x,y) - K(x,y)|
\label{a-a}
\end{equation}
\[
 \leqslant ||\psi||_2||\phi||_2 
    ||\sqrt{1-e^{-f}}||_2^2\,\sup_{x, y \in \supp f}
        |[Q_0G_LQ_0(1-Q_0G_LQ_0)^{-1}](x,y) - K(x,y)|,
\]
which tends to $0$, by Lemma \ref{B}(iv).
Thus the strong (in fact  the norm) convergence has been proved.
For the convergence of the trace norm, we use Lemma \ref{B}(iv) again and positive 
self-adjointness of operators $A_L$ and $K_f$ to get
\[
  ||A_L||_T - ||K_f||_T = \Tr[A_L - K_f] 
\]
\[
   = \int_{\R^d} dx\,(1-e^{-f(x)})
  ([Q_0G_LQ_0(1-Q_0G_LQ_0)^{-1}](x,x) - K(x,x))
            \to 0.
\] 
\hfill$\square$

\bigskip

By Lemma \ref{D} (iii, iv), we have
\[
    -\frac{\tilde z_0-z_0}{z_0}N+\frac{1- z_0}{z_0}N
    -\frac{1-\tilde z_0}{\tilde z_0}N =
    \frac{(\tilde z_0-z_0)(1-\tilde z_0)}{z_0\tilde
z_0}N=O\big(\frac{1}{N}\big).
\]
Applying Lemma \ref{C}(i) and Lemma \ref{F}(ii) to the righthand side of
\[
    -\frac{1- z_0}{z_0}p_0
    +\frac{1-\tilde z_0}{\tilde z_0}\tilde p_0
   = -\frac{1-\tilde g_0}{1-\tilde z_0\tilde g_0},
\]
we get the formula
\[
  {\rm E}_{L, N}^B\big[e^{-<f, \xi>}\big]=
\]
\begin{equation}
              \frac{\exp\big(-(\rho-\rho_c)(\sqrt{1-e^{-f}},
[1+W_L^*Q_0(1-Q_0G_LQ_0)^{-1}Q_0W_L]^{-1}\sqrt{1-e^{-f}})+o(1)\big)}
      { \Det [1+K_f]}.
\label{BEC_0}
\end{equation}
From the convergence $ W_L^*Q_0(1-Q_0G_LQ_0)^{-1}Q_0W_L = A_L \to K_f $ 
in the thermodynamic limit, we have proved the theorem.

\appendix

\section{ Complex integrals }
\begin{lem}
    For $ 0 \leqslant x \leqslant 1$ and $ p \geqslant 0$ satisfying
 $ 0 \leqslant px < 1 $, we have
\[
       1 \geqslant (1+x)^p(1-px) \geqslant
       \exp\Big(-\frac{p(1+p)(1+px^2)}{2(1-px)^2}x^2\Big).
\]
\end{lem}
{\sl Proof: } 
Put $ f(x) = \log(1+x)^p(1-px) $, then
\[
   f'(x) = \frac{p}{1+x} - \frac{p}{1-px}, \qquad
   f''(x) = - \frac{p(1+p)(1+px^2)}{(1+ x)^2(1-px)^2}
\]
hold.
So we have $ f(0)=0, \; f'(0) = 0 $ and
$ 0 \geqslant f''(\theta x) \geqslant - p(1+p)(1+px^2)/(1-px)^2$ for $\theta \in(0,1)$, 
which imply the result.
\hfill $\square$

\begin{lem}
Let the collection of numbers $\{\, p^{(N)}_j \, \}_{j,N}$
satisfies
\[
   p^{(N)}_0 >  p^{(N)}_1 \geqslant p^{(N)}_2 \geqslant \cdots
   \geqslant p^{(N)}_j \geqslant \cdots \geqslant 0, \quad
   \sum_{j=0}^{\infty}p^{(N)}_j = N.
\]

\medskip
\noindent Suppose that there exist a sequence $\, \{\, R^{(N)} \}_{N \in \N} $
and $ c \in (0,1) $ such that
\[
   1< R^{(N)}< cp_0^{(N)}\Big(1 \wedge \frac{1}{p_1^{(N)}} \Big),
\quad
   \lim_{N\to\infty}p_0^{(N)}/R^{(N)}e^{c'R^{(N)}} = 0
\]
and
\[
   \lim_{N\to\infty}\frac{R^{(N)2}\sum_{j=1}^{\infty}p_j^{(N)}
   (1+p_j^{(N)})}{p_0^{(N)2}} = 0, \qquad \mbox{ where } \quad 
    c' = \frac{\log(1+c)}{c}.
\]
Then
\[
   \lim_{N\to\infty}p_0^{(N)} \oint_{S_1(0)}\frac{d\eta}{2\pi i}
\frac{1}{\eta^{N+1}
         \prod_{j=0}^{\infty}(1 - p_j^{(N)}(\eta-1))} =
         \frac{1}{e}
\]
holds.
\end{lem}
{\sl Proof:} We omit the superscript $(N)$ here.

\medskip
\noindent Note that $ p_0 \to \infty$ and $ R \to \infty$ as $ N \to \infty$. 

\noindent By the preceding lemma,
\[
       1\geqslant \prod_{j=1}^{\infty}\Big[\Big(1+\frac{R}{p_0}\Big)^{p_j}
         \Big(1 - \frac{Rp_j}{p_0}\Big)\Big]
        \geqslant \exp\Big(-\sum_{j=1}^{\infty}\frac{p_j(1+p_j)}{2}
        \Big(1+\frac{R^2p_j}{p_0^2}\Big)\frac{R^2}{p_0^2}
         \Big( 1- \frac{Rp_j}{p_0}\Big)^{-2}\Big).
\]
So the assumption on $R$ implies
\[
     \prod_{j=1}^{\infty}\Big[\Big(1+\frac{R}{p_0}\Big)^{p_j}
         \Big(1 - \frac{Rp_j}{p_0}\Big)\Big]
         \underset{N\to\infty}{\longrightarrow} 1.
\]
Similarly, we have
\[
     \prod_{j=1}^{\infty}\Big[\Big(1+\frac{1}{p_0}\Big)^{p_j}
         \Big(1 - \frac{p_j}{p_0}\Big)\Big]
         \underset{N\to\infty}{\longrightarrow} 1.
\]

Now let us deform the integration contour of the complex variable $\eta$
to two parts
\[
      \oint_{S_1(0)} = - \oint_{S_{(R-1)/p_0}(1+1/p_0)} +
           \oint_{S_{1+R/p_0}(0)} = I_1 + I_2.
\]
$I_1$ is obtained by the residue at $ \eta = 1+1/p_0$:
\[
     I_1 = -p_0\Big[\Big(1+\frac{1}{p_0}\Big)^{N+1}(-p_0)\prod_{j=1}^{\infty}
          \Big(1-\frac{p_j}{p_0}\Big)\Big]^{-1}
\]
\[         = \Big(1+\frac{1}{p_0}\Big)^{-p_0-1 }\prod_{j=1}^{\infty}
          \Big[\Big(1+\frac{1}{p_0}\Big)^{p_j}
\Big(1-\frac{p_j}{p_0}\Big)\Big]^{-1}
           \underset{N\to\infty}{\longrightarrow} e^{-1}.
\]
$I_2$ can be estimated as
\[
     |I_2| \quad \leqslant \quad p_0 \int_{-\pi}^{\pi}\frac{d\theta}{2\pi}
    \prod_{j=0}^{\infty}\Big[\Big(1+\frac{R}{p_0}\Big)^{p_j}
   \Big|\,
1-p_j\bigg(\Big(1+\frac{R}{p_0}\Big)e^{i\theta}-1\bigg)\Big|\Big]^{-1}
\]
\[
    \leqslant p_0 \Big(1+\frac{R}{p_0}\Big)^{-p_0} |\, 1-R |^{-1}
      \Big[\prod_{j=1}^{\infty}\Big(1+\frac{R}{p_0}\Big)^{p_j}
   \Big(1-\frac{Rp_j}{p_0}\Big)\Big]^{-1}
     \underset{N\to\infty}{\longrightarrow} 0,
\]
since $(1 + R/p_0)^{p_0} \geqslant (1+c)^{R/c} =e^{c'R}$ and the assumption.
\hfill$\Box$

\end{document}